\documentclass[twocolumn, prl, superscriptaddress,notitlepage]{revtex4-2}
\usepackage{graphicx}
\usepackage{dcolumn}
\usepackage{bm}
\usepackage[usenames,dvipsnames]{color}
\usepackage[most]{tcolorbox}
\usepackage{multirow}
\usepackage{amssymb}
\usepackage{amsmath}
\usepackage{xcolor}
\usepackage[normalem]{ulem}

\usepackage{amsfonts}
\usepackage[colorlinks, linkcolor=blue,anchorcolor=blue,citecolor=blue,urlcolor=blue]{hyperref}

\usepackage[percent]{overpic}

\usepackage{pifont}
\usepackage{physics}
\usepackage{natbib}
\usepackage{xcolor}
\usepackage{subcaption}

\newcommand{\ii}{\mathrm{i}}

\newcommand{\equalcontrib}{These authors contributed equally.}
\usepackage{color,soul}

\begin{document}

\title{Control incompatibility in multiparameter quantum metrology}

\author{Zhiyao Hu }
\thanks{\equalcontrib}
\email{zhiyaohu.phys@gmail.com}
\affiliation{Pritzker School of Molecular Engineering, The University of Chicago, Chicago, Illinois 60637, USA}
\affiliation{
   MIT-Harvard Center for Ultracold Atoms and Research Laboratory of Electronics, Massachusetts Institute of Technology, Cambridge, MA 02139, USA}

\author{Shilin Wang}
\thanks{\equalcontrib}
\email{shilinnwang@gmail.com}
\affiliation{
   Department of Mechanical and Automation Engineering, The Chinese University of Hong Kong, Shatin, Hong Kong SAR, China}

\author{Linmu Qiao}
\thanks{\equalcontrib}
\affiliation{
   Department of Mechanical and Automation Engineering, The Chinese University of Hong Kong, Shatin, Hong Kong SAR, China}

  \author{Takuya Isogawa }
\affiliation{
   MIT-Harvard Center for Ultracold Atoms and Research Laboratory of Electronics, Massachusetts Institute of Technology, Cambridge, MA 02139, USA}
\affiliation{
   Department of Nuclear Science and Engineering, Massachusetts Institute of Technology, Cambridge, MA 02139, USA}
   
  \author{Changhao Li }
\affiliation{
   MIT-Harvard Center for Ultracold Atoms and Research Laboratory of Electronics, Massachusetts Institute of Technology, Cambridge, MA 02139, USA}
\affiliation{
   Department of Nuclear Science and Engineering, Massachusetts Institute of Technology, Cambridge, MA 02139, USA}

\author{Yu Yang}
\affiliation{
School of Physics, Xi'an Jiaotong University, Xi'an 710049, China
}

\author{Guoqing Wang}\email[]{gq\_wang@pku.edu.cn}
\affiliation{
   MIT-Harvard Center for Ultracold Atoms and Research Laboratory of Electronics, Massachusetts Institute of Technology, Cambridge, MA 02139, USA}
\affiliation{International Center for Quantum Materials, School of Physics, Peking University, Beijing 100871, China}

\author{Paola Cappellaro}\email[]{pcappell@mit.edu}
\affiliation{
   MIT-Harvard Center for Ultracold Atoms and Research Laboratory of Electronics, Massachusetts Institute of Technology, Cambridge, MA 02139, USA}
\affiliation{
   Department of Nuclear Science and Engineering, Massachusetts Institute of Technology, Cambridge, MA 02139, USA}
\affiliation{Department of Physics, Massachusetts Institute of Technology, Cambridge, MA 02139, USA}

\author{Haidong Yuan}\email[]{hdyuan@mae.cuhk.edu.hk}
\affiliation{
   Department of Mechanical and Automation Engineering, The Chinese University of Hong Kong, Shatin, Hong Kong SAR, China}
\affiliation{The Hong Kong Institute of Quantum Information Science and Technology, The Chinese University of Hong Kong, Shatin, Hong Kong SAR, China}
    \affiliation{State Key Laboratory of Quantum Information Technologies and Materials, The Chinese University of Hong Kong, Shatin, Hong Kong SAR, China}

\begin{abstract}
Quantum metrology strives to achieve the ultimate precision in parameter estimation. While fundamental limits for single-parameter estimation are well established, practical applications—such as quantum sensing, imaging, and field reconstruction—require the simultaneous estimation of multiple parameters. In such multiparameter settings, the attainable precision limits are far less understood, largely because the strategies that are optimal for estimating different parameters are generally incompatible. This incompatibility is especially pronounced in the design of optimal controls, where controls that enhance sensitivity to one parameter may degrade sensitivity to others. Here, we address the challenge of control incompatibility in multiparameter quantum estimation by introducing a geometric framework for constructing optimal controls that minimize incompatibility-induced trade-offs. Our approach clarifies the structure of these trade-offs and provides systematic tools for mitigating them. These findings establish a foundation for future research into control strategies that enable optimal estimation of multiple incompatible parameters, advancing the broader field of quantum metrology.
\end{abstract}

\maketitle

\paragraph{Introduction}---
A central pursuit in quantum metrology is determining the highest achievable precision and designing strategies to attain it. For single-parameter quantum estimation, significant progress has been made in understanding the ultimate limits of local precision limit and the corresponding optimal strategies \cite{introhelstrom1969quantum,introholevo2001statistical,introholevo2011probabilistic,qmMaccone_2004_review,qmGiovannetti_2011_review,qmGiovannetti_2006_metrology,qmdegen2017quantum,qmtoth2014quantum,qmpezze2018quantum,polino2020photonic,demille2024quantum,liu2024fully,liu2022optimal,lawrie2019quantum,pirandola2018advances,Fujiwara2008FibreBundle,Escher2011GeneralFramework,demkowicz2012elusive,kurdzialek2023using}. However, in practical applications, such as quantum sensing and quantum imaging, the estimation of multiple parameters is often necessary\cite{new2matsumoto2002new,new2kahn2009local,new2li2016fisher,new2pezze2017optimal,new2zhu2018universally,new2albarelli2019evaluating,new2yang2019optimal,new2yang2019attaining,new22005asymptotic,new22013quantum,new2sidhu2021tight,new2crowley2014tradeoff,new2vidrighin2014joint,new2yue2014quantum,new2zhang2014quantum,new2ragy2016compatibility,new2suzuki2016explicit,new2chen2017maximal,new2liu2017control,new2roccia2017entangling,new2chen2019optimal,new2razavian2020quantumness,new2sidhu2020geometric,new2candeloro2021properties,new2lu2021incorporating,new22005asymptotic,new2zhu2018universally,new2vidrighin2014joint,guhne2023colloquium,isogawa2026entanglement,isogawa2026approaching,hu2025optimal}, yet the ultimate precision limits in such scenarios remain far less understood. A major challenge in multi-parameter quantum metrology is the \textit{incompatibility}: 
strategies that are optimal for estimating different parameters can be mutually exclusive, preventing the simultaneous attainment of the single-parameter precision limits for all parameters. Understanding the trade-offs induced by the incompatibility is therefore a central problem in multiparameter quantum metrology \cite{new22005asymptotic,new22013quantum,new2albarelli2019evaluating,new2albarelli2020perspective,new2belliardo2021incompatibility,new2candeloro2021properties,new2carollo2019quantumness,new2chen2017maximal,new2chen2019optimal,new2conlon2021efficient,new2crowley2014tradeoff,new2demkowicz2020multi,new2hayashi2023tight,new2hayashi2024finding,new2hou2020minimal,new2hou2021super,new2hou2021zero,new2kahn2009local,new2li2016fisher,new2liu2017control,new2liu2020quantum,new2lu2021incorporating,new2matsumoto2002new,new2pezze2017optimal,new2ragy2016compatibility,new2razavian2020quantumness,new2roccia2017entangling,new2sidhu2020geometric,new2sidhu2021tight,new2suzuki2016explicit,new2szczykulska2016multi,new2vidrighin2014joint,new2yang2019attaining,new2yang2019optimal,new2yue2014quantum,new2zhang2014quantum,new2zhu2018universally,guhne2023colloquium,mpialbarelli2022probe,gorecki2022multiparameter,new2gill2000state,isogawa2026entanglement,isogawa2026approaching,hu2025optimal}.

Incompatibility can arise at every stage of a quantum metrology protocol, including the preparation of probe states, the implementation of controls during the evolution, and the execution of measurements. Trade-offs caused by the incompatibility of probe states and measurements have been extensively investigated~\cite{new22005asymptotic,new22013quantum,new2albarelli2019evaluating,new2albarelli2020perspective,new2belliardo2021incompatibility,new2candeloro2021properties,new2carollo2019quantumness,new2chen2017maximal,new2chen2019optimal,new2conlon2021efficient,new2crowley2014tradeoff,new2demkowicz2020multi,new2hayashi2023tight,new2hayashi2024finding,new2hou2020minimal,new2hou2021super,new2hou2021zero,new2kahn2009local,new2li2016fisher,new2liu2017control,new2liu2020quantum,new2lu2021incorporating,new2matsumoto2002new,new2pezze2017optimal,new2ragy2016compatibility,new2razavian2020quantumness,new2roccia2017entangling,new2sidhu2020geometric,new2sidhu2021tight,new2suzuki2016explicit,new2szczykulska2016multi,new2vidrighin2014joint,new2yang2019attaining,new2yang2019optimal,new2yue2014quantum,new2zhang2014quantum,new2zhu2018universally,guhne2023colloquium,mpialbarelli2022probe,gorecki2022multiparameter,new2gill2000state,new1chen2022information,mpichen2022incompatibility,xia2023toward}. By contrast, the trade-offs arising from \textit{control incompatibility} remain largely unexplored. Existing analytical bounds\cite{mpialbarelli2022probe} are typically not tight and provide limited guidance for designing optimal protocols. Numerical methods, such as gradient-based optimization\cite{new2liu2017control} and reinforcement learning\cite{XuReinforcement2021}, can find improved controls but are prone to local minima and offer limited physical insight into the nature of the trade-offs. The core difficulty lies in navigating the vast space of possible control strategies to identify those that optimally balance the competing demands of estimating multiple parameters simultaneously.

In this work, we introduce a geometric framework that addresses the challenge of control incompatibility. Central to our approach is a geometrical formulation that provides a clear and intuitive picture of the problem: we translate the search for optimal controls-which, in its original form, involves an intractable exploration over vast spaces of control sequences-into a transparent geometric optimization over the collective displacements of multiple vectors in the system's generator space. This formulation exposes the geometric origin of control-induced trade-offs and provides a systematic route to constructing explicit optimal controls that minimize them. Our framework therefore not only clarifies the structure of control incompatibility, but also offers practical tools for designing protocols that minimize the tradeoff induced by it.

We begin by presenting solutions for the paradigmatic problem of estimating parameters encoded in $SU(2)$ dynamics. In this setting, we show that a necessary condition for the absence of control incompatibility is that the angles between the generators remain constant in time. When this condition is violated-as is generically the case in multi-parameter estimation-we derive explicit optimal control protocols that minimize the resulting trade-off. Extending to $SU(N)$ dynamics, we derive an analytical bound on control incompatibility that imposes a fundamental precision limit on simultaneous estimation of multiple parameters. These results offer a unifying geometric picture and practical tools for understanding and reducing control incompatibility, advancing the performance of multiparameter quantum metrology.

For the estimation of multiple parameters \(\mathbf{x} = (x_1, \dots, x_n)\) encoded in a quantum state, the covariance matrix of any locally unbiased estimator \(\hat{\mathbf{x}}\) satisfies the Quantum Cramér-Rao Bound (QCRB)\cite{qcrbfisher1925theory,qcrbrao1992information,qmparis_quantum_2009,introhelstrom1969quantum,qcrbrao1992information,qmCaves_1994_Bures} with
$
\text{Cov}(\hat{\mathbf{x}}) \geq \frac{1}{\nu} J^{-1}(\mathbf{x}),
$
where \(\nu\) is the number of independent measurements and \(J(\mathbf{x})\) is the Quantum Fisher Information Matrix (QFIM)\cite{introhelstrom1969quantum}. 
For a pure state \(|\varphi_{\mathbf{x}}\rangle\), the elements of the QFIM are given by
$
J(\mathbf{x})_{ij} = 4 \, \text{Re} \left( \langle \partial_{x_i} \varphi_{\mathbf{x}} | \partial_{x_j} \varphi_{\mathbf{x}} \rangle - \langle \partial_{x_i} \varphi_{\mathbf{x}} | \varphi_{\mathbf{x}} \rangle \langle \varphi_{\mathbf{x}} | \partial_{x_j} \varphi_{\mathbf{x}} \rangle \right).
$
For a single parameter \(x_i\), the bound simplifies to \(\delta x_i^2 \geq (\nu J(\mathbf{x})_{ii})^{-1}\), or equivalently $\frac{1}{\delta x_i^2}\leq \nu J(\mathbf{x})_{ii}$. Summing across all parameters gives the collective bound $\sum_i \frac{1}{\delta x_i^2} \leq \nu \sum_i J(x)_{ii}$. 

A convenient way to compute the QFI is through the generator associated with each parameter. 
For a unitary evolution \(|\varphi_{\mathbf{x}}\rangle = U_{\mathbf{x}} |\varphi_0\rangle\), the generator for $x_i$, written in the Heisenberg picture, is
\begin{equation}\label{eq:generator}
S_{x_i}(T) = \int_0^T U_{\mathbf{x}}^{\dagger}(t) \left[ \partial_{x_i} H(\mathbf{x}, t) \right] U_{\mathbf{x}}(t)  dt.    
\end{equation}
The QFI for $x_i$ is directly proportional to the variance of this generator $
J(\mathbf{x})_{ii} = 4 \, \text{Var}(S_{x_i}(T))=4(\langle \varphi_0|S_{x_i}^2(T)|\varphi_0\rangle-\langle \varphi_0|S_{x_i}(T)|\varphi_0\rangle^2)$. Maximizing the precision for a parameter is therefore equivalent to maximizing the variance of its generator. By selecting the optimal initial probe state and optimal control for a single parameter $x_i$, one achieves the maximal QFI $J_{x_i}^{max}=(\int_0^T|\partial_{x_i} H(\mathbf{x}, t)|dt)^2$, where $|\cdot|$ denotes the semi-norm given by the difference between the maximal and minimal eigenvalues. This bounds $\frac{1}{\delta \hat{x}_i^2}$ as $\frac{1}{\delta \hat{x}_i^2}\leq \nu J_{x_i}^{max}$. 

In multi-parameter scenarios, summing the individually optimized single-parameter bounds gives the trivial upper bound
 $\sum_i \frac{1}{\delta \hat{x}_i^2}\leq \nu \sum_i J_{x_i}^{max}$. This bound, however, is generally not attainable because the control strategies for achieving each individual $J_{x_i}^{max}$ are typically incompatible with one another. For any single strategy used to estimate all parameters simultaneously, we have $\sum_i \frac{1}{\delta \hat{x}_i^2}\leq \nu \sum_i J(x)_{ii}\leq \nu \sum_i J_{x_i}^{max}$. The gap between $\sum_i J(x)_{ii}$, achievable by one common strategy, and $\sum_i J_{x_i}^{max}$, obtained by optimizing each parameter separately, quantifies the tradeoff induced by the incompatibility of the optimal strategies for different parameters. 
In this paper, we study this tradeoff induced from the incompatibility of the optimal control for different parameters. We develop geometric methods to derive tighter bounds on the maximum $Tr(J)$ achievable with any single strategy. For the case of estimating two parameters under SU(2) dynamics, we identify an optimal control protocol that attains the maximal $Tr(J)$. 

We now specialize to $SU(2)$ dynamics. A general controlled Hamiltonian can be written as
$
H(\mathbf{x},t) = H_0(\mathbf{x},t)+H_C(t)
$,
where $H_0(\mathbf{x},t)=\vec{F}(\mathbf{x},t) \cdot \vec{\sigma}$ is the parameter-dependent free Hamiltonian with \(\vec{\sigma} = (\sigma_x, \sigma_y, \sigma_z)\) as the Pauli matrices, and $H_C(t)$ is the control Hamiltonian that can be externally tuned. In the supplementary material, we show that for SU(2) dynamics, the maximally entangled state between the probe qubit and an ancilla, \(|\varphi_0\rangle = \frac{1}{\sqrt{2}}(|00\rangle + |11\rangle)\), is universally optimal. This is because the reduced state, \(\mathrm{Tr}_A[|\varphi_0\rangle\langle\varphi_0|] = I/2\), maximizes the variance for any generator \(S_{x_i}(T) = \vec{s}_{x_i}(T) \cdot \vec{\sigma}\). With this optimal probe state, the quantum Fisher information for $x_i$ simplifies to
$
J(\mathbf{x})_{ii} = 4 \cdot \frac{1}{2} \mathrm{Tr}(S_{x_i}^2(T)) = 4 \|\vec{s}_{x_i}(T)\|^2,
$
where $\|\vec{v}\|^2=v_1^2+v_2^2+v_3^2$ is the Euclidean norm. 

This yields a simple geometric picture. For the generator in Eq.(\ref{eq:generator}), we can rewrite the parameter derivative of the Hamiltonian as \(\partial_{x_i} H(\mathbf{x},t) = \vec{v}_{x_i}(t) \cdot \vec{\sigma}\), where $\vec{v}_{x_i}(t)$ can be viewed as an instantaneous velocity vector. Under the isomorphism between $SU(2)$ and $SO(3)$,  conjugation by \(U_{\mathbf{x}}(t)\) corresponds to a rotation \(R(t) \in SO(3)\) acting on this velocity vector. Consequently, the generator can be represented as a net displacement vector \(S_{x_i}(T) = \vec{s}_{x_i}(T) \cdot \vec{\sigma}\) with
$
\vec{s}_{x_i}(T) = \int_0^T R(t)  \vec{v}_{x_i}(t)  dt.
$
Thus maximizing the QFI for $x_i$ is equivalent to choosing a rotation trajectory \(R(t)\) that maximizes the total length of the displacement vector \(\vec{s}_{x_i}(T)\). For a single parameter $x_i$, the optimal strategy is to rotate all instantaneous velocities along one fixed direction. This makes the trajectory in generator space a straight line and yields the maximal QFI\cite{qmpang_optimal_2017,new2hou2021zero,new2hou2021super}
\begin{equation}\label{eq:maxQFI}
J_{x_i}^{max} =4 \left( \int_0^T \| \vec{v}_{x_i}(t) \|  dt \right)^2.
\end{equation} 
For multiple parameters, however, the rotation trajectory \( R(t) \) that aligns the velocity vectors for one parameter generally fails to align those for another. This creates a tradeoff induced by the control incompatibility. We quantify this trade-off by defining the control incompatibility measure $
\mathcal{G} = \sum_i J^{max}_{x_i} - \mathrm{Tr}[J(\mathbf{x})].$ The first term is the sum of the individually achievable single-parameter optima, while the second term is the total QFI achieved by a single common control protocol. Thus 
\(\mathcal{G}\) measures the loss of the total QFI because one control strategy must serve all parameters simultaneously. 
A smaller \( \mathcal{G} \) indicates weaker incompatibility and \( \mathcal{G} = 0 \) means that a single control protocol can simultaneously achieve the maximal QFI for all parameters. This definition can be generalized to a weighted form, \( \mathcal{G}(\mathbf{w}) = \sum_i w_i (J^{max}_{x_i} - J(\mathbf{x})_{ii}) \), to prioritize the estimation of specific parameters. Since each $J^{max}_{x_i}$ is known from single-parameter optimization with \(J^{max}_{x_i}=4 \left( \int_0^T \| \vec{v}_{x_i}(t) \|  dt \right)^2\), the key challenge reduces to maximizing $Tr[J(x)]$ under a single control strategy.

\begin{figure}[htbp]
\includegraphics[width=0.5\textwidth]{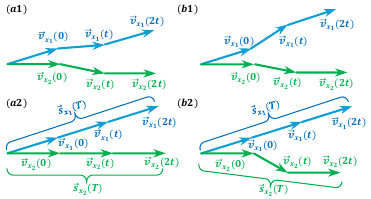}
\caption{
Geometric illustration of control (in)compatibility. (a) Compatible case: The angle \(\theta_{ij}(t)\) between the velocity vectors \(\vec{v}_{x_i}(t)\) and \(\vec{v}_{x_j}(t)\) remains constant over time, i.e., \(\theta_{ij}(0) = \theta_{ij}(t) = \theta_{ij}(2t)\). In this scenario, the optimal control that aligns the generator for \(x_i\) simultaneously aligns the generator for \(x_j\), allowing both parameters to achieve their maximal QFI simultaneously. (b) Incompatible case: The angle \(\theta_{ij}(t)\) varies with time, such that $\theta_{ij}(0)$, $\theta_{ij}(t)$ and $\theta_{ij}(2t)$ are different. Here, no single control can perfectly align both generators; the optimal control that maximizes the QFI for \(x_i\) inevitably leaves the generator for \(x_j\) misaligned, giving rise to an unavoidable trade-off in estimation precision.
}
\label{ci}
\end{figure}

We first establish a simple geometric necessary condition for the absence of control incompatibility: the instantaneous angle between every pair of velocity vectors remains constant throughout the entire evolution. That is,
\begin{equation}
    \theta_{ij}(t) = \theta_{ij}(0), 
    \qquad 
    \forall\, t \in [0,T],
    \label{nec}
\end{equation}
where 
    $\cos\theta_{ij}(t)
    =
    \frac{
        \vec{v}_{x_i}(t)\cdot\vec{v}_{x_j}(t)
    }{
        \|\vec{v}_{x_i}(t)\|\,\|\vec{v}_{x_j}(t)\|
    }.$
As illustrated in Fig.~\ref{ci}(a), a single control can simultaneously maximize the displacement vectors  $\vec{s}_{x_i}(T)$ and  $\vec{s}_{x_j}(T)$ only if the angle between their instantaneous velocities, $\vec{v}_{x_i}(t)$ and  $\vec{v}_{x_j}(t)$, remains constant. In this scenario, if a rotation $R(t)$ aligns $R(t)\vec{v}_{x_i}(t)$ along a fixed axis, the other rotated velocity vector $R(t)\vec{v}_{x_j}(t)$ automatically aligned along another fixed axis separated by the constant angle $\theta_{ij}$. 

A particularly important example is the time- independent free Hamiltonian, $H_0(x, t) = H_0(x)$. In this case, this condition is inherently satisfied. Because each derivative \(\partial_{x_i} H(\mathbf{x})\) is time independent, so the direction of every velocity vector \(\vec{v}_{x_i}\) is fixed, and the angles between all pairs of velocity vectors are automatically constant. The optimal rotation can then be chosen as $R(t)=I$. This choice preserves the direction of each velocity vector-\(R(t)\vec{v}_{x_i} = \vec{v}_{x_i}\) for all \(x_i\)—and therefore yields straight-line trajectories for all parameters. Physically, this can be implemented by a control Hamiltonian that cancels the free Hamiltonian, \(H_C(t) = -H(\mathbf{x})\). Since the true values of the parameters are unknown, in practice this is impelmented adaptively as \(H_C(t) = -H(\hat{\mathbf{x}})\), where \(\hat{\mathbf{x}}\) s the current best estimate obtained from prior knowledge and accumulated measurement data. 
  
Conversely, if $\theta_{ij}(t)$ varies with time, control incompatibility becomes unavoidable. Since a global rotation 
$R(t)$ intrinsically preserves the instantaneous angles between vectors, simultaneous alignment with fixed target axes is impossible unless the relative angles between the velocity vectors exactly match the angles between the aligned axes. Thus, a time-dependent $\theta_{ij}$ prevents simultaneous alignment at all times, as illustrated in Fig.~\ref{ci}(b). In this incompatible regime, the goal becomes finding the best compromise between competing alignment requirements. Specifically, we seek the optimal control rotation \( R(t) \in SO(3) \) that maximizes \( \mathrm{Tr}[J(\mathbf{x})] \), thereby minimizing $\mathcal{G}$. 

\begin{figure}[hb]
\includegraphics[width=0.5\textwidth]{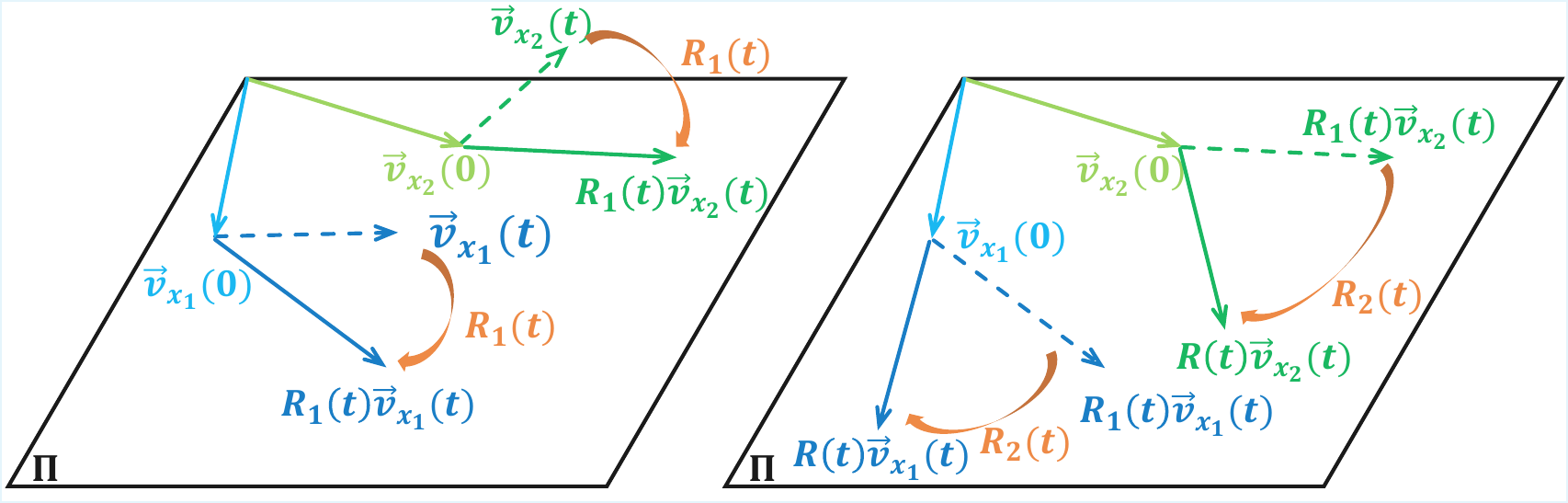}
\caption{
Geometric illustration of optimal control protocol: 
The first control step, \(R_1(t)\), rotates both vectors into a fixed plane \(\Pi = \mathrm{span}\{\vec v_{x_1}(0),\vec v_{x_2}(0)\}\), eliminating out-of-plane distortions. The second control step, \(R_2(t)\), then performs an in-plane rotation that minimizes the angular separation between the two vectors, thereby reducing control incompatibility to its minimal level. 
}
\label{fig:R12}
\end{figure}

For two-parameter estimation, the necessary condition in Eq.(\ref{nec}) is also sufficient. The optimal control that aligns the generator for one parameter can simultaneously align the generator for the other parameter, as illustrated in Fig. \ref{ci}(a). More importantly, for the general case where control incompatibility is present (i.e., when \(\theta_{ij}(t)\) fluctuates), the optimal control can still be derived explicitly. In this situation, the optimal control rotation can be decomposed into two geometrically distinct components
 \(   R(t)=R_{2}(t)R_{1}(t), \)
as illustrated in Fig.~\ref{fig:R12}. The first component, $R_{1}(t)$, rotates $\vec{v}_{x_1}(t)$ and $\vec{v}_{x_2}(t)$ into a fixed plane. Choosing this plane to be the $X-Y$ plane, for example, yields $R_{1}(t)\vec{v}_{x_1}(t)=(a_{x_1}(t), b_{x_1}(t), 0)^T$ and $R_{1}(t)\vec{v}_{x_2}(t)=(a_{x_2}(t), b_{x_2}(t), 0)^T$. This rotation removes all out-of-plane distortions so that the remaining incompatibility arises purely from in-plane angular variations (see supplementary material for detailed proof). The second component, $R_2(t)$, performs an in-plane rotation that maximizes the net displacement of $R_{1}(t)\vec{v}_{x_1}(t)$ and $R_{1}(t)\vec{v}_{x_2}(t)$ in the plane . When the fixed plane is chosen as the $X-Y$ plane, $R_2(t)$ is a rotation about the $Z$-axis with an optimal angle $\alpha(t)$. To make the optimization concrete, we represent the in-plane vectors as complex numbers. By letting
$z_k(t) = a_{x_k}(t) + i\, b_{x_k}(t)$
be the complex representation of \(R_1(t)\vec{v}_{x_k}(t)\) with $k=1, 2$, the 
$Z$-axis rotation becomes a multiplication by the phase factor \(e^{i\alpha(t)}\). The resulting net displacements are given by
$S_k = \int_0^T e^{i\alpha(t)} z_k(t)\, dt,$ $k = 1, 2.$  Our objective is now to find the optimal \(\alpha(t)\) that maximizes \(|S_1|^2 + |S_2|^2\). 

Defining the complex vector $\vec{S} = (S_1, S_2)^T$, its norm $\|\vec{S}\|=\sqrt{|S_1|^2+|S_2|^2}$ is equal to its maximum projection onto a unit vector $\vec{C} = (C_1, C_2)^T$:
\begin{equation}
\aligned
    \|\vec{S}\| &= \max_{\|\vec{C}\|=1} \text{Re} \left( C_1 S_1 + C_2 S_2 \right) \\
    &= \max_{\|\vec{C}\|=1} \text{Re} \int_0^T e^{\ii\alpha(t)} \big( C_1 z_1(t) + C_2 z_2(t) \big) \dd t.
    \endaligned
\end{equation}
Interchanging the order of maximization yields
\begin{equation}
    \max_{\alpha(t)} \|\vec{S}\| = \max_{\|\vec{C}\|=1} \left[ \max_{\alpha(t)} \int_0^T \text{Re} \Big( e^{\ii\alpha(t)} W_C(t) \Big) \dd t \right],
\end{equation}
where $W_C(t) = \sum_{k=1}^2 C_k z_k(t)$. For any fixed $\vec{C}$, the inner integral is maximized pointwise by choosing $\alpha(t) = -\arg(W_C(t))$, so that $\text{Re} \Big( e^{\ii\alpha(t)} W_C(t) \Big)=|W_C(t)|$, here $W_C(t)= e^{\ii \arg(W_C(t))}|W_C(t)|$. The functional optimization over $\alpha(t)$ thus reduces to a finite-dimensional algebraic optimization over $\vec{C}=(C_1,C_2)^T$ with
\begin{equation}
    \max_{\alpha(t)} \|\vec{S}\| = \max_{\|\vec{C}\|=1} \int_0^T \left|C_1 z_1(t)+C_2z_2(t) \right| \dd t.
\end{equation}
By parameterizing the unit vector as $C_1=\sqrt{p}e^{\ii \theta_1}$ and $C_2=\sqrt{1-p}e^{\ii\theta_2}$ with $p\in[0,1]$ and $\theta_1,\theta_2\in R$, the problem reduces to optimizing just three real parameters: \(p\), \(\theta_1\), and \(\theta_2\). This transforms the original problem from a functional optimization over the infinite-dimensional space of time-dependent angles \(\alpha(t)\) to a tractable, finite-dimensional algebraic optimization, making the optimal control both analytically transparent and computationally tractable. 

Finally for the estimation of an arbitrary number of parameters, this framework provides an analytical upper bound on $Tr[J(x)]$.
Since \( J(\mathbf{x})_{ii} = 4 \|\vec{s}_{x_i}(T)\|^2 \), we have
$
\mathrm{Tr}[J(\mathbf{x})] = \sum_i 4 \left\| \int_0^T R(t) \vec{v}_{x_i}(t)  dt \right\|^2$. Expanding this expression into a double integral gives:
\begin{eqnarray}
\aligned
    \mathrm{Tr}[J(\mathbf{x})] &= 4 \sum_i \int_0^T\!\!\int_0^T \left[ \vec{v}_{x_i}^T(t_1) R^T(t_1) R(t_2) \vec{v}_{x_i}(t_2) \right] dt_1 dt_2\\
    &=4 \int_0^T\!\!\int_0^T \mathrm{Tr}\left[ R^T(t_1) R(t_2)  \mathcal{V}(t_1, t_2) \right] dt_1 dt_2,
\endaligned
\end{eqnarray}
where we define the matrix \( \mathcal{V}(t_1, t_2) = \sum_i \vec{v}_{x_i}(t_1) \vec{v}_{x_i}^T(t_2) \). By applying the inequality \(\mathrm{Tr}\left[ R^T(t_1) R(t_2)  \mathcal{V}(t_1, t_2) \right] \leq |\mathcal{V}(t_1, t_2)|_*\), where \(|\cdot|_*\) denotes the trace norm that equals the sum of singular values, we obtain an analytical upper bound  
\begin{equation}\label{eq:bound}
\mathrm{Tr}[J(\mathbf{x})] \leq 4\int_0^T\int_0^T |\mathcal{V}(t_1, t_2)|_* \, dt_1 dt_2.
\end{equation}

This upper bound naturally generalizes to the estimation of multiple parameters encoded in general SU(\(N\)) dynamics. In this setting, a general Hamiltonian can be written as
$
H(\mathbf{x},t) = H_0(\mathbf{x},t)+H_C(t),
$
with the free Hamiltonian given by \(H_0(\mathbf{x},t)=\vec{F}(\mathbf{x},t) \cdot \vec{\Lambda}\), here \(\vec{\Lambda}=(\Lambda_1,\cdots, \Lambda_{N^2-1})\) form an orthonormal basis for \(N\)-dimensional Hermitian matrices. We can again express both the parameter generator and the Hamiltonian derivative in this basis as \(S_{x_i}(T) = \vec{s}_{x_i}(T) \cdot \vec{\Lambda}\) and \(\partial_{x_i}H(\mathbf{x},t) = \vec{v}_{x_i}(t) \cdot \vec{\Lambda}\). These vectors satisfy the geometric relation 
$
\vec{s}_{x_i}(T) = \int_0^T R(t)  \vec{v}_{x_i}(t)  \, dt,
$
where the rotation now resides in a higher-dimensional space, \(R(t) \in \mathrm{SO}(N^2-1)\). Proceeding analogously, we arrive at the bound 
$
\mathrm{Tr}[J(\mathbf{x})] \leq 4\int_0^T\int_0^T |\mathcal{V}(t_1, t_2)|_* \, dt_1 dt_2
$
with \(
\mathcal{V}(t_1, t_2) = \sum_{i} \vec{v}_{x_i}(t_1) \vec{v}^T_{x_i}(t_2)
\)(see supplementary material for detailed derivation)


In the End Matter, we present several paradigmatic examples illustrating the optimal control scheme. We begin with the fully compatible regime, in which control incompatibility is absent, and show that our geometric framework naturally recovers known results in the literature. We then turn to the incompatible regime, where no single control can simultaneously optimize all parameters, and demonstrate how our approach quantifies and characterizes the resulting unavoidable trade-offs. 

\paragraph{Summary}-We have developed a geometric framework for addressing the largely unexplored problem of control incompatibility in multiparameter quantum estimation. For SU(2) dynamics, we derive the necessary and sufficient condition for the complete absence of control incompatibility. More importantly, when incompatibility is unavoidable, our framework yields explicit control protocols that minimize the resulting trade-offs. Crucially, this geometric formulation transforms the search for optimal controls—traditionally a computationally intractable functional problem—into a highly tractable algebraic optimization over just a few parameters. These results fill an important gap in the theory of multiparameter quantum metrology. 

A natural next step is to extend the present framework to noisy dynamics. One promising route is the purification-based approach\cite{Fujiwara2008FibreBundle,Escher2011GeneralFramework,kurdzialek2023using,demkowicz2012elusive,liu2022optimal}, in which noisy parameter-dependent channels are represented as unitary evolutions on an enlarged system--environment Hilbert space. This formulation may enable the geometric description of control incompatibility to be generalized from closed-system generators to effective purified generators for noisy dynamics. Another direction is to quantify control incompatibility using alternative objective functions, such as weighted estimation covariance. Together with ongoing developments in understanding measurement and probe incompatibility, these extensions could provide a more comprehensive characterization of the ultimate precision limits in multiparameter quantum metrology.

\acknowledgments
We thank Boning Li and Minh-Thi Nguyen for the helpful discussion.
H.Y. acknowledges partial support from the Quantum Science and Technology-National Science and Technology Major Project (2023ZD0300600), the Research Grants Council of Hong Kong (14309223, 14309624, 14309022), Guangdong Provincial Quantum Science Strategic Initiative (GDZX2303007, GDZX2505003) and 1+1+1 CUHK-CUHK(SZ)-GDST Joint Collaboration Fund (Grant No. GRDP2025-022). Guoqing Wang acknowledges support by The Fundamental Research Funds for the Central Universities, Peking University. Yu Yang acknowledges support by the Fundamental Research Funds for the Central Universities (Grant No. xxj032025044).

\bibliography{main_text} 

@article{isogawa2026entanglement,
  title={Entanglement-assisted multiparameter estimation with a solid-state quantum sensor},
  author={Isogawa, Takuya and Wang, Guoqing and Li, Boning and Hu, Zhiyao and Nishimura, Shunsuke and Kanamoto, Ayumi and Yuan, Haidong and Cappellaro, Paola},
  journal={PRX Quantum},
  volume={7},
  number={2},
  pages={020307},
  year={2026},
  publisher={APS}
}

@article{isogawa2026approaching,
  title={Approaching the Limit in Multiparameter AC Magnetometry with Quantum Control},
  author={Isogawa, Takuya and Hu, Zhiyao and Kanamoto, Ayumi and Phadetsuwannukun, Nutdech and Wang, Shilin and Nishimura, Shunsuke and Li, Boning and Jiang, Liang and Saleem, Zain H and Wang, Guoqing and others},
  journal={arXiv preprint arXiv:2602.17648},
  year={2026}
}

@article{hu2025optimal,
  title={Optimal scheme for distributed quantum metrology},
  author={Hu, Zhiyao and Zang, Allen and Wang, Jianwei and Zhong, Tian and Yuan, Haidong and Jiang, Liang and Saleem, Zain H},
  journal={arXiv preprint arXiv:2509.18334},
  year={2025}
}

@article{introhelstrom1969quantum,
  title={Quantum detection and estimation theory},
  author={Helstrom, Carl W},
  journal={Journal of Statistical Physics},
  volume={1},
  pages={231--252},
  year={1969},
  publisher={Springer}
}

@book{introholevo2011probabilistic,
  title={Probabilistic and statistical aspects of quantum theory},
  author={Holevo, Alexander S},
  volume={1},
  year={2011},
  publisher={Springer Science \& Business Media}
}

@book{introholevo2001statistical,
  title={Statistical structure of quantum theory},
  author={Holevo, Alexander S},
  year={2001},
  publisher={Springer Science \& Business Media}
}

@article{XuReinforcement2021,
  title = {Generalizable control for multiparameter quantum metrology},
  author = {Xu, Han and Wang, Lingna and Yuan, Haidong and Wang, Xin},
  journal = {Phys. Rev. A},
  volume = {103},
  issue = {4},
  pages = {042615},
  numpages = {13},
  year = {2021},
  month = {Apr},
  publisher = {American Physical Society},
}

@article{new1chen2022information,
  title={Information geometry under hierarchical quantum measurement},
  author={Chen, Hongzhen and Chen, Yu and Yuan, Haidong},
  journal={Physical Review Letters},
  volume={128},
  number={25},
  pages={250502},
  year={2022},
  publisher={APS}
}

@article{new1yuan2017quantum,
  title={Quantum parameter estimation with general dynamics},
  author={Yuan, Haidong and Fung, Chi-Hang Fred},
  journal={npj Quantum Information},
  volume={3},
  number={1},
  pages={14},
  year={2017},
  publisher={Nature Publishing Group UK London}
}

@article{new2gill2000state,
  title={State estimation for large ensembles},
  author={Gill, Richard D and Massar, Serge},
  journal={Physical Review A},
  volume={61},
  number={4},
  pages={042312},
  year={2000},
  publisher={APS}
}

@article{new2albarelli2020perspective,
  title={A perspective on multiparameter quantum metrology: From theoretical tools to applications in quantum imaging},
  author={Albarelli, Francesco and Barbieri, Marco and Genoni, Marco G and Gianani, Ilaria},
  journal={Physics Letters A},
  volume={384},
  number={12},
  pages={126311},
  year={2020},
  publisher={Elsevier}
}

@article{new2carollo2019quantumness,
  title={On quantumness in multi-parameter quantum estimation},
  author={Carollo, Angelo and Spagnolo, Bernardo and Dubkov, Alexander A and Valenti, Davide},
  journal={Journal of Statistical Mechanics: Theory and Experiment},
  volume={2019},
  number={9},
  pages={094010},
  year={2019},
  publisher={IOP Publishing}
}

@article{new2lu2021incorporating,
  title={Incorporating Heisenberg’s uncertainty principle into quantum multiparameter estimation},
  author={Lu, Xiao-Ming and Wang, Xiaoguang},
  journal={Physical Review Letters},
  volume={126},
  number={12},
  pages={120503},
  year={2021},
  publisher={APS}
}

@article{new2ragy2016compatibility,
  title={Compatibility in multiparameter quantum metrology},
  author={Ragy, Sammy and Jarzyna, Marcin and Demkowicz-Dobrza{\'n}ski, Rafa{\l}},
  journal={Physical Review A},
  volume={94},
  number={5},
  pages={052108},
  year={2016},
  publisher={APS}
}

@article{new2chen2017maximal,
  title={Maximal quantum Fisher information matrix},
  author={Chen, Yu and Yuan, Haidong},
  journal={New Journal of Physics},
  volume={19},
  number={6},
  pages={063023},
  year={2017},
  publisher={IOP Publishing}
}

@article{new2chen2019optimal,
  title={Optimal joint estimation of multiple Rabi frequencies},
  author={Chen, Hongzhen and Yuan, Haidong},
  journal={Physical Review A},
  volume={99},
  number={3},
  pages={032122},
  year={2019},
  publisher={APS}
}

@article{new2liu2020quantum,
  title={Quantum Fisher information matrix and multiparameter estimation},
  author={Liu, Jing and Yuan, Haidong and Lu, Xiao-Ming and Wang, Xiaoguang},
  journal={Journal of Physics A: Mathematical and Theoretical},
  volume={53},
  number={2},
  pages={023001},
  year={2020},
  publisher={IOP Publishing}
}

@article{new2demkowicz2020multi,
  title={Multi-parameter estimation beyond quantum Fisher information},
  author={Demkowicz-Dobrza{\'n}ski, Rafa{\l} and G{\'o}recki, Wojciech and Gu{\c{t}}{\u{a}}, M{\u{a}}d{\u{a}}lin},
  journal={Journal of Physics A: Mathematical and Theoretical},
  volume={53},
  number={36},
  pages={363001},
  year={2020},
  publisher={IOP Publishing}
}

@article{new2sidhu2020geometric,
  title={Geometric perspective on quantum parameter estimation},
  author={Sidhu, Jasminder S and Kok, Pieter},
  journal={AVS Quantum Science},
  volume={2},
  number={1},
  year={2020},
  publisher={AIP Publishing}
}

@article{new2vidrighin2014joint,
  title={Joint estimation of phase and phase diffusion for quantum metrology},
  author={Vidrighin, Mihai D and Donati, Gaia and Genoni, Marco G and Jin, Xian-Min and Kolthammer, W Steven and Kim, MS and Datta, Animesh and Barbieri, Marco and Walmsley, Ian A},
  journal={Nature communications},
  volume={5},
  number={1},
  pages={3532},
  year={2014},
  publisher={Nature Publishing Group UK London}
}

@article{new2crowley2014tradeoff,
  title={Tradeoff in simultaneous quantum-limited phase and loss estimation in interferometry},
  author={Crowley, Philip JD and Datta, Animesh and Barbieri, Marco and Walmsley, Ian A},
  journal={Physical Review A},
  volume={89},
  number={2},
  pages={023845},
  year={2014},
  publisher={APS}
}

@article{new2yue2014quantum,
  title={Quantum-enhanced metrology for multiple phase estimation with noise},
  author={Yue, Jie-Dong and Zhang, Yu-Ran and Fan, Heng},
  journal={Scientific reports},
  volume={4},
  number={1},
  pages={5933},
  year={2014},
  publisher={Nature Publishing Group UK London}
}

@article{new2zhang2014quantum,
  title={Quantum metrological bounds for vector parameters},
  author={Zhang, Yu-Ran and Fan, Heng},
  journal={Physical Review A},
  volume={90},
  number={4},
  pages={043818},
  year={2014},
  publisher={APS}
}

@article{new2liu2017control,
  title={Control-enhanced multiparameter quantum estimation},
  author={Liu, Jing and Yuan, Haidong},
  journal={Physical Review A},
  volume={96},
  number={4},
  pages={042114},
  year={2017},
  publisher={APS}
}

@article{new2roccia2017entangling,
  title={Entangling measurements for multiparameter estimation with two qubits},
  author={Roccia, Emanuele and Gianani, Ilaria and Mancino, Luca and Sbroscia, Marco and Somma, Fabrizia and Genoni, Marco G and Barbieri, Marco},
  journal={Quantum Science and Technology},
  volume={3},
  number={1},
  pages={01LT01},
  year={2017},
  publisher={IOP Publishing}
}

@article{new2razavian2020quantumness,
  title={On the quantumness of multiparameter estimation problems for qubit systems},
  author={Razavian, Sholeh and Paris, Matteo GA and Genoni, Marco G},
  journal={Entropy},
  volume={22},
  number={11},
  pages={1197},
  year={2020},
  publisher={MDPI}
}

@article{new2candeloro2021properties,
  title={On the properties of the asymptotic incompatibility measure in multiparameter quantum estimation},
  author={Candeloro, Alessandro and Paris, Matteo GA and Genoni, Marco G},
  journal={Journal of Physics A: Mathematical and Theoretical},
  volume={54},
  number={48},
  pages={485301},
  year={2021},
  publisher={IOP Publishing}
}

@article{new2matsumoto2002new,
  title={A new approach to the Cram{\'e}r-Rao-type bound of the pure-state model},
  author={Matsumoto, Keiji},
  journal={Journal of Physics A: Mathematical and General},
  volume={35},
  number={13},
  pages={3111},
  year={2002},
  publisher={IOP Publishing}
}

@article{new22013quantum,
  title={Quantum enhanced multiple phase estimation},
  author={Humphreys, Peter C and Barbieri, Marco and Datta, Animesh and Walmsley, Ian A},
  journal={Physical review letters},
  volume={111},
  number={7},
  pages={070403},
  year={2013},
  publisher={APS}
}

@article{new2pezze2017optimal,
  title={Optimal measurements for simultaneous quantum estimation of multiple phases},
  author={Pezz{\`e}, Luca and Ciampini, Mario A and Spagnolo, Nicol{\`o} and Humphreys, Peter C and Datta, Animesh and Walmsley, Ian A and Barbieri, Marco and Sciarrino, Fabio and Smerzi, Augusto},
  journal={Physical review letters},
  volume={119},
  number={13},
  pages={130504},
  year={2017},
  publisher={APS}
}

@book{new22005asymptotic,
  title={Asymptotic theory of quantum statistical inference: selected papers},
  author={Hayashi, Masahito},
  year={2005},
  publisher={World Scientific}
}

@article{new2hou2020minimal,
  title={Minimal tradeoff and ultimate precision limit of multiparameter quantum magnetometry under the parallel scheme},
  author={Hou, Zhibo and Zhang, Zhao and Xiang, Guo-Yong and Li, Chuan-Feng and Guo, Guang-Can and Chen, Hongzhen and Liu, Liqiang and Yuan, Haidong},
  journal={Physical Review Letters},
  volume={125},
  number={2},
  pages={020501},
  year={2020},
  publisher={APS}
}

@article{new2belliardo2021incompatibility,
  title={Incompatibility in quantum parameter estimation},
  author={Belliardo, Federico and Giovannetti, Vittorio},
  journal={New Journal of Physics},
  volume={23},
  number={6},
  pages={063055},
  year={2021},
  publisher={IOP Publishing}
}

@article{new2hou2021zero,
  title={Zero--trade-off multiparameter quantum estimation via simultaneously saturating multiple Heisenberg uncertainty relations},
  author={Hou, Zhibo and Tang, Jun-Feng and Chen, Hongzhen and Yuan, Haidong and Xiang, Guo-Yong and Li, Chuan-Feng and Guo, Guang-Can},
  journal={Science Advances},
  volume={7},
  number={1},
  pages={eabd2986},
  year={2021},
  publisher={American Association for the Advancement of Science}
}

@article{new2hou2021super,
  title={“Super-Heisenberg” and Heisenberg scalings achieved simultaneously in the estimation of a rotating field},
  author={Hou, Zhibo and Jin, Yan and Chen, Hongzhen and Tang, Jun-Feng and Huang, Chang-Jiang and Yuan, Haidong and Xiang, Guo-Yong and Li, Chuan-Feng and Guo, Guang-Can},
  journal={Physical Review Letters},
  volume={126},
  number={7},
  pages={070503},
  year={2021},
  publisher={APS}
}

@article{new2yang2019optimal,
  title={Optimal measurements for quantum multiparameter estimation with general states},
  author={Yang, Jing and Pang, Shengshi and Zhou, Yiyu and Jordan, Andrew N},
  journal={Physical Review A},
  volume={100},
  number={3},
  pages={032104},
  year={2019},
  publisher={APS}
}

@article{new2li2016fisher,
  title={Fisher-symmetric informationally complete measurements for pure states},
  author={Li, Nan and Ferrie, Christopher and Gross, Jonathan A and Kalev, Amir and Caves, Carlton M},
  journal={Physical review letters},
  volume={116},
  number={18},
  pages={180402},
  year={2016},
  publisher={APS}
}

@article{new2zhu2018universally,
  title={Universally Fisher-symmetric informationally complete measurements},
  author={Zhu, Huangjun and Hayashi, Masahito},
  journal={Physical review letters},
  volume={120},
  number={3},
  pages={030404},
  year={2018},
  publisher={APS}
}

@article{new2szczykulska2016multi,
  title={Multi-parameter quantum metrology},
  author={Szczykulska, Magdalena and Baumgratz, Tillmann and Datta, Animesh},
  journal={Advances in Physics: X},
  volume={1},
  number={4},
  pages={621--639},
  year={2016},
  publisher={Taylor \& Francis}
}

@article{new2albarelli2019evaluating,
  title={Evaluating the Holevo Cram{\'e}r-Rao bound for multiparameter quantum metrology},
  author={Albarelli, Francesco and Friel, Jamie F and Datta, Animesh},
  journal={Physical review letters},
  volume={123},
  number={20},
  pages={200503},
  year={2019},
  publisher={APS}
}

@article{new2kahn2009local,
  title={Local asymptotic normality for finite dimensional quantum systems},
  author={Kahn, Jonas and Gu{\c{t}}{\u{a}}, M{\u{a}}d{\u{a}}lin},
  journal={Communications in Mathematical Physics},
  volume={289},
  number={2},
  pages={597--652},
  year={2009},
  publisher={Springer}
}

@article{new2yang2019attaining,
  title={Attaining the ultimate precision limit in quantum state estimation},
  author={Yang, Yuxiang and Chiribella, Giulio and Hayashi, Masahito},
  journal={Communications in Mathematical Physics},
  volume={368},
  pages={223--293},
  year={2019},
  publisher={Springer}
}

@article{new2suzuki2016explicit,
  title={Explicit formula for the Holevo bound for two-parameter qubit-state estimation problem},
  author={Suzuki, Jun},
  journal={Journal of Mathematical Physics},
  volume={57},
  number={4},
  year={2016},
  publisher={AIP Publishing}
}

@article{new2sidhu2021tight,
  title={Tight bounds on the simultaneous estimation of incompatible parameters},
  author={Sidhu, Jasminder S and Ouyang, Yingkai and Campbell, Earl T and Kok, Pieter},
  journal={Physical Review X},
  volume={11},
  number={1},
  pages={011028},
  year={2021},
  publisher={APS}
}

@article{new2conlon2021efficient,
  title={Efficient computation of the Nagaoka--Hayashi bound for multiparameter estimation with separable measurements},
  author={Conlon, Lorc{\'a}n O and Suzuki, Jun and Lam, Ping Koy and Assad, Syed M},
  journal={npj Quantum Information},
  volume={7},
  number={1},
  pages={110},
  year={2021},
  publisher={Nature Publishing Group UK London}
}

@article{new2hayashi2023tight,
  title={Tight Cram{\'e}r-Rao type bounds for multiparameter quantum metrology through conic programming},
  author={Hayashi, Masahito and Ouyang, Yingkai},
  journal={Quantum},
  volume={7},
  pages={1094},
  year={2023},
  publisher={Verein zur F{\"o}rderung des Open Access Publizierens in den Quantenwissenschaften}
}

@article{new2hayashi2024finding,
  title={Finding the optimal probe state for multiparameter quantum metrology using conic programming},
  author={Hayashi, Masahito and Ouyang, Yingkai},
  journal={npj Quantum Information},
  volume={10},
  number={1},
  pages={111},
  year={2024},
  publisher={Nature Publishing Group UK London}
}

@article{qmparis_quantum_2009,
  author = {Paris, Matteo G. A.},
title = {QUANTUM ESTIMATION FOR QUANTUM TECHNOLOGY},
journal = {International Journal of Quantum Information},
volume = {07},
number = {supp01},
pages = {125-137},
year = {2009}
}

@article{qmCaves_1994_Bures,
  title = {Statistical distance and the geometry of quantum states},
  author = {Braunstein, Samuel L. and Caves, Carlton M.},
  journal = {Phys. Rev. Lett.},
  volume = {72},
  issue = {22},
  pages = {3439--3443},
  numpages = {0},
  year = {1994},
  month = {May},
  publisher = {American Physical Society},
 
}

@article{qmMaccone_2004_review,
author = {Vittorio Giovannetti  and Seth Lloyd  and Lorenzo Maccone },
title = {Quantum-Enhanced Measurements: Beating the Standard Quantum Limit},
journal = {Science},
volume = {306},
number = {5700},
pages = {1330-1336},
year = {2004}}

@article{qmGiovannetti_2011_review,
  title={Advances in quantum metrology},
  author={Giovannetti, Vittorio and Lloyd, Seth and Maccone, Lorenzo},
  journal={Nature photonics},
  volume={5},
  number={4},
  pages={222--229},
  year={2011},

  publisher={Nature Publishing Group UK London}
}

@article{qmGiovannetti_2006_metrology,
  title={Quantum metrology},
  author={Giovannetti, Vittorio and Lloyd, Seth and Maccone, Lorenzo},
  journal={Physical review letters},
  volume={96},
  number={1},
  pages={010401},
  year={2006},
   
  publisher={APS}
}

@article{qmtoth2014quantum,
  title={Quantum metrology from a quantum information science perspective},
  author={T{\'o}th, G{\'e}za and Apellaniz, Iagoba},
  journal={Journal of Physics A: Mathematical and Theoretical},
  volume={47},
  number={42},
  pages={424006},
  year={2014},
  
  publisher={IOP Publishing}
}

@article{qmpezze2018quantum,
  title={Quantum metrology with nonclassical states of atomic ensembles},
  author={Pezz{\`e}, Luca and Smerzi, Augusto and Oberthaler, Markus K and Schmied, Roman and Treutlein, Philipp},
  journal={Reviews of Modern Physics},
  volume={90},
  number={3},
  pages={035005},
  year={2018},
 
  publisher={APS}
}

@article{qmdegen2017quantum,
  title={Quantum sensing},
  author={Degen, Christian L and Reinhard, Friedemann and Cappellaro, Paola},
  journal={Reviews of modern physics},
  volume={89},
  number={3},
  pages={035002},
  year={2017},
  publisher={APS}
}

@article{qmpang_optimal_2017,
  title = {Optimal Adaptive Control for Quantum Metrology with Time-Dependent {{Hamiltonians}}},
  author = {Pang, Shengshi and Jordan, Andrew N.},
  year = {2017},
  month = apr,
  journal = {Nature Communications},
  volume = {8},
  number = {1},
  pages = {14695},
  issn = {2041-1723},
 
  urldate = {2020-09-25},
  langid = {english}
}

@inproceedings{qcrbfisher1925theory,
  title={Theory of statistical estimation},
  author={Fisher, Ronald Aylmer},
  booktitle={Mathematical proceedings of the Cambridge philosophical society},
  volume={22},
  number={5},
  pages={700--725},
  year={1925},
 
  organization={Cambridge University Press}
}

@article{qcrbrao1992information,
  title={Information and the accuracy attainable in the estimation of statistical parameters},
  author={Rao, C Radhakrishna and others},
  journal={Breakthroughs in statistics},
  pages={235--247},
  year={1992}
}

@article{mpyuan_sequential_2016,
  title = {Sequential {{Feedback Scheme Outperforms}} the {{Parallel Scheme}} for {{Hamiltonian Parameter Estimation}}},
  author = {Yuan, Haidong},
  year = {2016},
  month = oct,
  journal = {Physical Review Letters},
  volume = {117},
  number = {16},
  pages = {160801},
  issn = {0031-9007, 1079-7114},

  langid = {english}
}

@article{mpialbarelli2022probe,
  title={Probe incompatibility in multiparameter noisy quantum metrology},
  author={Albarelli, Francesco and Demkowicz-Dobrza{\'n}ski, Rafa{\l}},
  journal={Physical Review X},
  volume={12},
  number={1},
  pages={011039},
  year={2022},

  publisher={APS}
}

@article{gorecki2022multiparameter,
  title={Multiparameter quantum metrology in the Heisenberg limit regime: Many-repetition scenario versus full optimization},
  author={G{\'o}recki, Wojciech and Demkowicz-Dobrza{\'n}ski, Rafa{\l}},
  journal={Physical Review A},
  volume={106},
  number={1},
  pages={012424},
  year={2022},
  publisher={APS}
}

@article{kurdzialek2023using,
  title={Using adaptiveness and causal superpositions against noise in quantum metrology},
  author={Kurdzia{\l}ek, Stanis{\l}aw and G{\'o}recki, Wojciech and Albarelli, Francesco and Demkowicz-Dobrza{\'n}ski, Rafa{\l}},
  journal={Physical Review Letters},
  volume={131},
  number={9},
  pages={090801},
  year={2023},
  publisher={APS}
}

@article{polino2020photonic,
  title={Photonic quantum metrology},
  author={Polino, Emanuele and Valeri, Mauro and Spagnolo, Nicol{\`o} and Sciarrino, Fabio},
  journal={AVS Quantum Science},
  volume={2},
  number={2},
  year={2020},
  publisher={AIP Publishing}
}

@article{mpichen2022incompatibility,
  title={Incompatibility measures in multiparameter quantum estimation under hierarchical quantum measurements},
  author={Chen, Hongzhen and Chen, Yu and Yuan, Haidong},
  journal={Physical Review A},
  volume={105},
  number={6},
  pages={062442},
  year={2022},

  publisher={APS}
}

@article{xia2023toward,
  title={Toward incompatible quantum limits on multiparameter estimation},
  author={Xia, Binke and Huang, Jingzheng and Li, Hongjing and Wang, Han and Zeng, Guihua},
  journal={Nature Communications},
  volume={14},
  number={1},
  pages={1021},
  year={2023},
  publisher={Nature Publishing Group UK London}
}

@article{guhne2023colloquium,
  title={Colloquium: Incompatible measurements in quantum information science},
  author={G{\"u}hne, Otfried and Haapasalo, Erkka and Kraft, Tristan and Pellonp{\"a}{\"a}, Juha-Pekka and Uola, Roope},
  journal={Reviews of Modern Physics},
  volume={95},
  number={1},
  pages={011003},
  year={2023},
  publisher={APS}
}

@article{demille2024quantum,
  title={Quantum sensing and metrology for fundamental physics with molecules},
  author={DeMille, David and Hutzler, Nicholas R and Rey, Ana Maria and Zelevinsky, Tanya},
  journal={Nature Physics},
  pages={1--9},
  year={2024},
  publisher={Nature Publishing Group UK London}
}

@article{liu2024fully,
  title={Fully-Optimized Quantum Metrology: Framework, Tools, and Applications},
  author={Liu, Qiushi and Hu, Zihao and Yuan, Haidong and Yang, Yuxiang},
  journal={Advanced Quantum Technologies},
  pages={2400094},
  year={2024},
  publisher={Wiley Online Library}
}

@article{liu2022optimal,
  title={Optimal scheme for quantum metrology},
  author={Liu, Jing and Zhang, Mao and Chen, Hongzhen and Wang, Lingna and Yuan, Haidong},
  journal={Advanced Quantum Technologies},
  volume={5},
  number={1},
  pages={2100080},
  year={2022},
  publisher={Wiley Online Library}
}

@article{pirandola2018advances,
  title={Advances in photonic quantum sensing},
  author={Pirandola, Stefano and Bardhan, B Roy and Gehring, Tobias and Weedbrook, Christian and Lloyd, Seth},
  journal={Nature Photonics},
  volume={12},
  number={12},
  pages={724--733},
  year={2018},
  publisher={Nature Publishing Group UK London}
}

@article{lawrie2019quantum,
  title={Quantum sensing with squeezed light},
  author={Lawrie, Benjamin J and Lett, Paul D and Marino, Alberto M and Pooser, Raphael C},
  journal={Acs Photonics},
  volume={6},
  number={6},
  pages={1307--1318},
  year={2019},
  publisher={ACS Publications}
}

@article{demkowicz2012elusive,
  title={The elusive Heisenberg limit in quantum-enhanced metrology},
  author={Demkowicz-Dobrza{\'n}ski, Rafa{\l} and Ko{\l}ody{\'n}ski, Jan and Gu{\c{t}}{\u{a}}, M{\u{a}}d{\u{a}}lin},
  journal={Nature communications},
  volume={3},
  number={1},
  pages={1063},
  year={2012},
  publisher={Nature Publishing Group UK London}
}

@article{Khaneja2005GRAPE,
  author  = {Khaneja, Navin and Reiss, Timo and Kehlet, Cindie and Schulte-Herbr{\"u}ggen, Thomas and Glaser, Steffen J.},
  title   = {Optimal control of coupled spin dynamics: Design of {NMR} pulse sequences by gradient ascent algorithms},
  journal = {Journal of Magnetic Resonance},
  volume  = {172},
  number  = {2},
  pages   = {296--305},
  year    = {2005}
}

@article{PhysRevA.96.012117,
  title = {Quantum parameter estimation with optimal control},
  author = {Liu, Jing and Yuan, Haidong},
  journal = {Phys. Rev. A},
  volume = {96},
  issue = {1},
  pages = {012117},
  numpages = {14},
  year = {2017},
  month = {Jul},
  publisher = {American Physical Society}
}

@article{Fujiwara2008FibreBundle,
  author  = {Fujiwara, Akio and Imai, Hiroshi},
  title   = {A Fibre Bundle over Manifolds of Quantum Channels and Its
             Application to Quantum Statistics},
  journal = {J. Phys. A: Math. Theor.},
  volume  = {41},
  number  = {25},
  pages   = {255304},
  year    = {2008}
}

@article{Escher2011GeneralFramework,
  author  = {Escher, B. M. and de Matos Filho, R. L. and Davidovich, L.},
  title   = {General Framework for Estimating the Ultimate Precision Limit
             in Noisy Quantum-Enhanced Metrology},
  journal = {Nat. Phys.},
  volume  = {7},
  number  = {5},
  pages   = {406--411},
  year    = {2011}
}
\appendix
\clearpage
\section*{End Matter}
To illustrate our optimal-control scheme, we consider several paradigmatic examples. The first two consider scenarios where control incompatibility is absent and a globally optimal control exists: the first involves a time-independent free Hamiltonian, while the second features a time-dependent free Hamiltonian. Although these problems have previously been analyzed using methods tailored to their specific structures~\cite{new1yuan2017quantum,new2hou2021super}, we show that they emerge naturally as special cases of our general framework, with the optimal controls obtained directly from our general protocol. The third example, in contrast, addresses a situation where control incompatibility is present, going beyond the scope of prior work.


\textbf{Example 1: Time-independent free Hamiltonian}. Consider the estimation of all three components of a DC magnetic field, described by the time-independent Hamiltonian $ H_0(B_x,B_y, B_z)=B_x\sigma_x+B_y\sigma_y+B_z\sigma_z$. The velocity vectors for the three parameters are 
$\vec v_{B_x}(t)=(1,0,0)^{T}$, $\vec v_{B_y}(t)=(0,1,0)^{T}$ and $\vec v_{B_z}(t)=(0,0,1)^{T}$, which are constant in time and mutually orthogonal. Hence the angles between all pairs of velocity vectors remain fixed throughout the evolution, and the system lies in the fully compatible regime. The generators are already aligned in fixed directions, so the optimal control rotation may be chosen as $ R(t)=I$. Physically, this rotation can be implemented by cancelling the free Hamiltonian with an adaptive control, $H_C(t)=-H_0(\hat{B})$ with $\hat{B}$ as the estimated value of $B=(B_x,B_y, B_z)$. Under this optimal control, each generator accumulates along a straight trajectory of length $T$, and the QFI for each parameter attains its single-parameter maximum: $J_{B_i}=4T^2$, $\forall i=x,y,z$. This recovers the known result that the optimal control for estimating time-independent Hamiltonian parameters can be taken as $H_c(t)=-H_0(\hat{x})$~\cite{mpyuan_sequential_2016,new2hou2021zero}.The analytical bound in Eq.~\eqref{eq:bound} is also tight in this example. Since $\mathcal{V}(t_1, t_2)=\sum_{i=x,y,z} \vec{v}_{B_i}(t_1) \vec{v}_{B_i}^T(t_2)=I$, Eq.(\ref{eq:bound}) yields 
\(J_{B_x}+J_{B_y}+J_{B_z}\leq 12 T^2\), which is saturated by the optimal control above.

\textbf{Example 2: Time-dependent free Hamiltonian without control incompatibility}. We next consider the simultaneous estimation of the amplitude and frequency of an AC magnetic field, described by the time-dependent Hamiltonian $H_0(B,\omega,t)=-B(\cos \omega t \sigma_x+\sin \omega t \sigma_z) $\cite{qmpang_optimal_2017,new2hou2021super}. The corresponding velocity vectors are $\vec v_{B}(t)=(-\cos\omega t,\,0,\,-\sin\omega t)^{T}$ and $\vec v_{\omega}(t)
= B t\,(\sin\omega t,\,0,\,-\cos\omega t)^{T}$. Since $\vec v_{B}(t)\cdot \vec v_{\omega}(t)=0$, the angle $\theta_{B\omega}(t)=\frac{\pi}{2}$ remains constant for all $t$. Consequently, control incompatibility is absent, and the maximal QFI for $B$ and $\omega$ can then be achieved simultaneously. 

Because both velocity vectors are already in the $X-Z$ plane, we can let $R_1(t)=I$. For the in-plane rotation, $R_2(t)=e^{\,\alpha(t)\, Y}$ with $Y=\begin{pmatrix}
0&0&1\\
0&0&0\\
-1&0&0
\end{pmatrix}$
 being the generator for the rotation around Y-axis in SO(3). Representing the in-plane vectors as complex numbers with $z_1(t)=-cos(\omega t)-\ii \sin(\omega t)=-e^{\ii \omega t}$, $z_2(t)=Bt[\sin(\omega t)-\ii \cos(\omega t)]=-\ii Bte^{\ii \omega t}$, we have
 \begin{equation}
 \aligned
    \max_{\alpha(t)} \|\vec{S}\| &= \max_{\|\vec{C}\|=1} \int_0^T \left|C_1 z_1(t)+C_2z_2(t) \right| \dd t\\
    & =\max_{p,\theta_1,\theta_2} \int_0^T \left|-\sqrt{p}e^{\ii \theta_1} e^{\ii \omega t}-\sqrt{1-p}e^{\ii \theta_2}iBte^{\ii\omega t} \right| \dd t\\
   & \leq \max_{p} \int_0^T \sqrt{p}+\sqrt{1-p}Bt dt\\
   &\leq \sqrt{T^2+\frac{1}{4}B^2T^4}, 
\endaligned
\end{equation}
where the first inequality is saturated with $\theta_2-\theta_1=-\frac{\pi}{2}$ and the last inequality is saturated with $p^\star=\frac{4}{4+B^2T^2}$. The optimal $\alpha(t)$ that saturate the bound is given by $\alpha(t)=-\arg(W_C(t))$ with $W_C(t)=-\sqrt{p^\star} e^{\ii \theta_{1}}e^{\ii\omega t}-\sqrt{1-p^\star}e^{\ii \theta_{2}}iBte^{\ii\omega t}$. Using the phase condition \(\theta_2-\theta_1=-\pi/2\), this becomes
 \begin{equation}
 \aligned
\alpha(t)&=-arg[-\sqrt{p^\star}e^{\ii \theta_{1}}e^{\ii\omega t}-\sqrt{1-p^\star}e^{\ii \theta_{2}}iBte^{\ii\omega t}]\\
&=-arg[e^{\ii (\omega t+\theta_1+\pi)}(\sqrt{p^\star}+\sqrt{1-p^\star}Bt)]\\
&=-\omega t-\theta_1-\pi.
\endaligned
\end{equation}
Since an overall constant phase shift does not affect the alignment, we can arbitrarily choose $\theta_1=-\pi$, simplifying the control angle to $\alpha(t)=-\omega t$. Physically, this rotation is achieved by applying the control Hamiltonian $H_C(t)=-H_0-\frac{\dot{\alpha}(t)}{2}\sigma_y=-H_0-\frac{\omega}{2}\sigma_y$. This yields the maximal $Tr(J)=4\max_{\alpha(t)} \|\vec{S}\|^2=4T^2+B^2T^4$, in agreement with previous studies ~\cite{qmpang_optimal_2017,new2hou2021super}. The analytical upper bound in Eq.~\eqref{eq:bound} is also tight for this model. Direct evaluation gives  $4
    \int_0^T\!\!\int_0^T
    \|\mathcal V(t_1,t_2)\|_*\,dt_1dt_2
    = 4T^2+B^2T^4$, which is exactly saturated by the optimal control above.

\begin{figure}[htbp]
    \centering

    \begin{subfigure}{0.5\textwidth}

        \phantomsubcaption
        \label{fig:suba}

        \phantomsubcaption
        \label{fig:subb}

        \phantomsubcaption
        \label{fig:subc}

        \hspace*{-0.8cm}
        \begin{overpic}[width=\columnwidth]{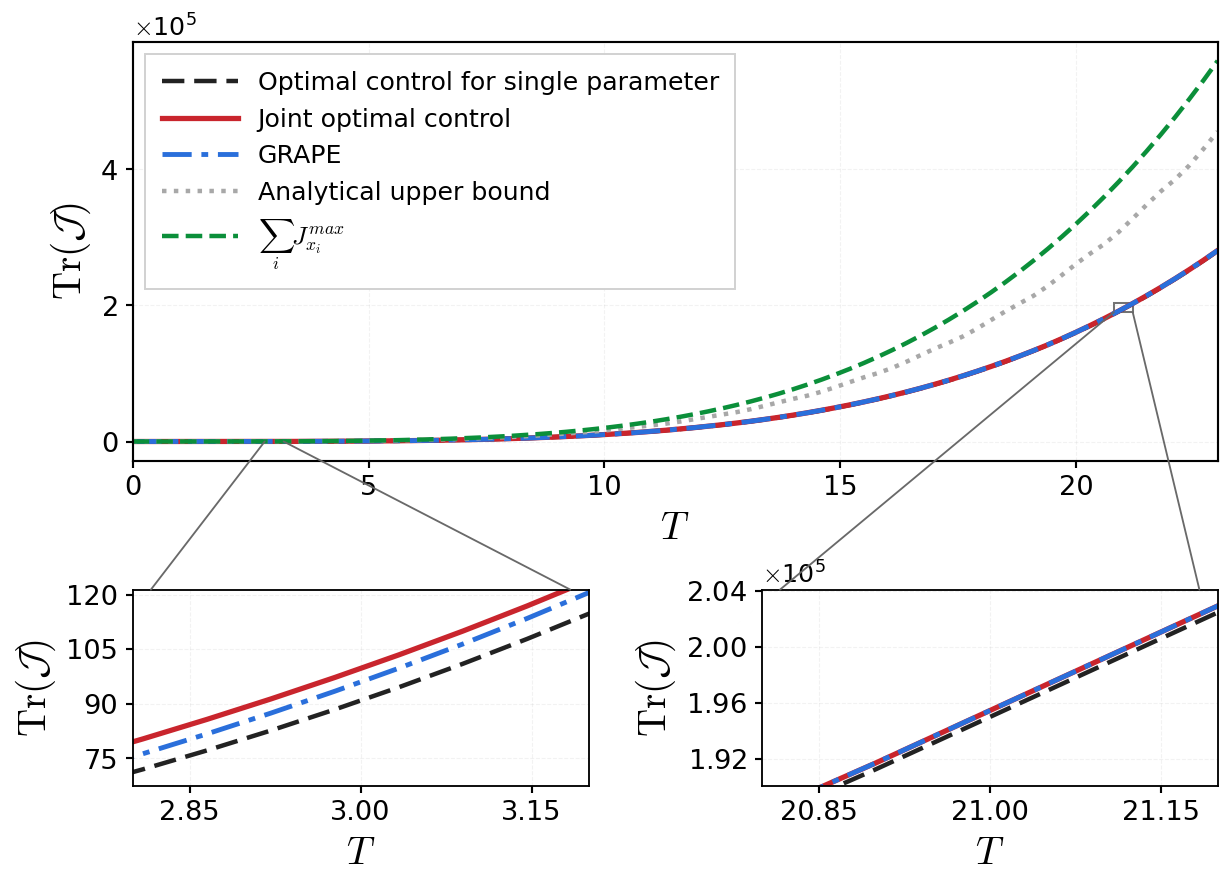}

            \put(6,70){%
                \makebox(0,0)[lt]{%
                    \fontfamily{ptm}\selectfont
                    \bfseries\itshape\footnotesize
                    (a)%
                }%
            }

            \put(6,29){%
                \makebox(0,0)[lt]{%
                    \fontfamily{ptm}\selectfont
                    \bfseries\itshape\footnotesize
                    (b)%
                }%
            }

            \put(57,29){%
                \makebox(0,0)[lt]{%
                    \fontfamily{ptm}\selectfont
                    \bfseries\itshape\footnotesize
                    (c)%
                }%
            }

        \end{overpic}
    \end{subfigure}

    \begin{subfigure}[b]{\columnwidth}

        \phantomsubcaption
        \label{fig:subd}

        \phantomsubcaption
        \label{fig:sube}

        \begin{overpic}[width=0.49\columnwidth]{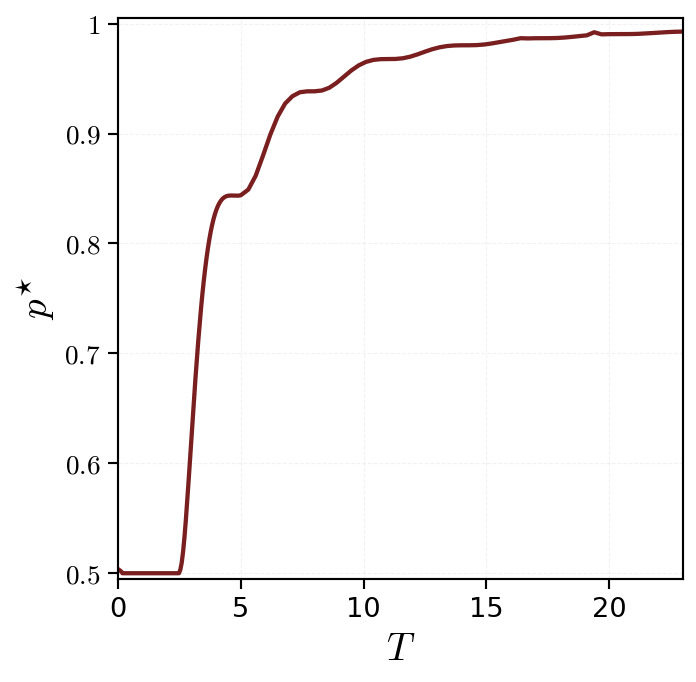}
            \put(5,106){%
                \makebox(0,0)[lt]{%
                    \fontfamily{ptm}\selectfont
                    \bfseries\itshape\footnotesize
                    (d)%
                }%
            }
        \end{overpic}
        \hfill
        \begin{overpic}[width=0.47\columnwidth]{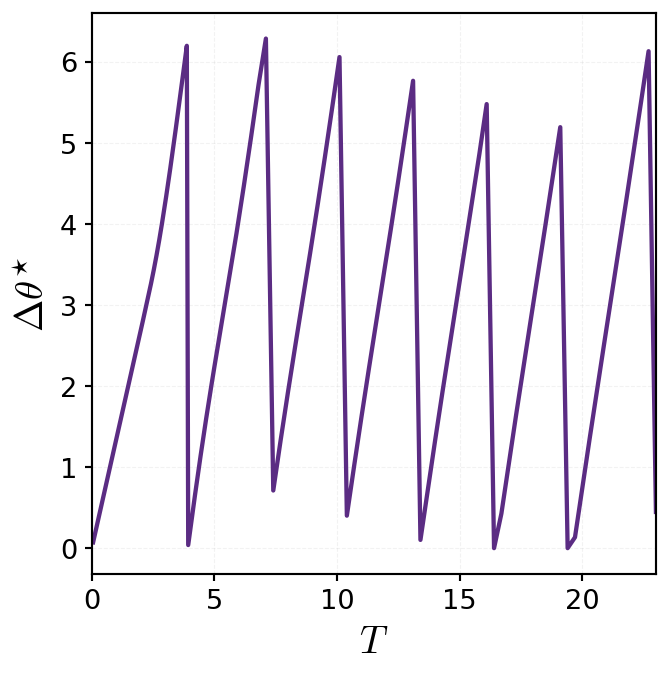}
            \put(6,107){%
                \makebox(0,0)[lt]{%
                    \fontfamily{ptm}\selectfont
                    \bfseries\itshape\footnotesize
                    (e)%
                }%
            }
        \end{overpic}

    \end{subfigure}

    \caption{
               (a) Comparison of different strategies for Example~3, including
        1: optimal control for a single parameter, where the control is
        taken as the optimal control for the estimation of a single
        parameter $\omega_1$ with
        $\alpha(t)=-\omega_1t$;
        2: joint optimal control, where the control is taken as
        Eq.~(\ref{eq:optimalalpha});
        3: GRAPE, where the control is obtained numerically using
        Gradient Ascent Pulse Engineering
        \cite{Khaneja2005GRAPE,PhysRevA.96.012117,new2liu2017control};
        4: analytical upper bound, which is obtained from
        Eq.~(\ref{eq:bound});
        5: $\sum_i J_{x_i}^{\max}$, where $J_{x_i}^{\max}$ is the
        maximal achievable QFI when only $x_i$ is unknown.
        (b) and (c) Two zoomed regions in the short- and long-time
        regimes are also included. Here $B_1=B_2=1$, $\omega_1=1$,
        and $\omega_2=-1$.
        (d) and (e) The optimal $p^\star$ and
        $\Delta\theta^\star$ with respect to $T$.
    }
    \label{fig:TrJ}

\end{figure}

\textbf{Example 3: Time-dependent free Hamiltonian with control incompatibility.} We now consider an example demonstrating control incompatibility, extending beyond the scenarios of previous studies.

Consider the simultaneous estimation of two frequencies of an AC magnetic field, governed by the Hamiltonian 
$H_0(\omega,t)=B[(\cos \omega_1 t+\cos \omega_2 t )\sigma_x +  (\sin \omega_1t+\sin \omega_2 t)\sigma_z]$. The corresponding velocity vectors are 
$\vec v_{\omega_1}(t)=B(-t\sin \omega_1 t,\,0,\,t\cos \omega_1 t)^{T}$ 
and 
$\vec v_{\omega_2}(t)=B(-t\sin \omega_2 t,\,0,\,t\cos \omega_2 t)^{T}$. A key observation is that the relative orientation of these vectors changes over time. 
Their instantaneous angle is
\begin{equation}
\cos\theta_{12}(t)
=
\frac{\vec v_{\omega_1}(t)\!\cdot\!\vec v_{\omega_2}(t)}
{\|\vec v_{\omega_1}(t)\|\,\|\vec v_{\omega_2}(t)\|}
=\cos[(\omega_1-\omega_2)t],
\end{equation}
which explicitly depends on $t$. This time-dependence leads to inherent control incompatibility: no single global rotation can align both vectors simultaneously at all times. 

Because both vectors already lie in the $X-Z$ plane, the first-stage rotation is trivial($R_1(t)=I$). 
Representing the in-plane vectors as complex numbers, $z_1(t) = i Bt e^{\ii \omega_1 t}, \quad z_2(t) = iB t e^{\ii \omega_2 t}$. Under the in-plane rotation, $R_2(t)=e^{\,\alpha(t)\, Y}$, the generators of the parameters are given by $S_k = \int_0^T e^{\ii\alpha(t)} z_k(t) \dd t$. The optimization then proceeds as
\begin{equation}
\aligned
    \max_{\alpha(t)} \|\vec{S}\| &= \max_{|C_1|^2+|C_2|^2=1} \int_0^T B\left| C_1 \ii t e^{\ii \omega_1 t} + C_2 \ii  t e^{\ii \omega_2 t} \right| \dd t\\
    &=\max_{p,\Delta \theta} \int_0^T Bt\left| \sqrt{p}    + \sqrt{1-p}   e^{\ii (\Delta \omega t+\Delta \theta)} \right| \dd t\\
  \endaligned  
\end{equation}
where $\Delta \omega=\omega_2-\omega_1$ and $\Delta\theta=\theta_2-\theta_1$. The optimal parameters $p^\star$ and $\Delta \theta^\star$ depend on the evolution time $T$, as shown in Fig.(\ref{fig:subd}) and Fig.(\ref{fig:sube}). The optimal $\alpha(t)$ can be explicitly obtained as
\begin{equation}\label{eq:optimalalpha}
\alpha(t)=-\omega_1 t-
\operatorname{atan}\frac{
\sqrt{1-p^\star}\sin(\Delta\omega t+\Delta\theta^\star)}{
\sqrt{p^\star}
+
\sqrt{1-p^\star}\cos(\Delta\omega t+\Delta\theta^\star)},
\end{equation}
which can be implemented via the control Hamiltonian $H_C(t)=-H_0-\frac{\dot{\alpha}(t)}{2}\sigma_y$. As shown in Fig.(\ref{fig:subd}), at the short time regime, $p^\star=\frac{1}{2}$. In this regime the control angle simplifies to $\alpha(t)=-\omega_1 t-\frac{\Delta \omega t+\Delta \theta^\star}{2}=-\frac{\omega_1+\omega_2}{2}t-\frac{\Delta \theta^\star}{2}$, which can be taken as $\alpha(t)=-\frac{\omega_1+\omega_2}{2}t$ by ignoring the constant term. At the long time regime, $p^\star\rightarrow 1$, $\alpha(t)\rightarrow-\omega_1 t$, which converges to the optimal control for the estimation of a single frequency. 
\end{document}